\documentclass[12pt,showpacs,amssymb]{revtex4}
\usepackage{mathrsfs}

\begin{document}

\title{Canonical Realizations of Doubly
Special Relativity}

\author{Pablo
\surname{Gal\'an}}\email{galan@iem.cfmac.csic.es}

\author{Guillermo A. \surname{Mena Marug\'an}}
\email{mena@iem.cfmac.csic.es} \affiliation{Instituto
de Estructura de la Materia, CSIC, Serrano 121-123,
28006 Madrid, Spain}

\begin{abstract}

Doubly Special Relativity is usually formulated in
momentum space, providing the explicit nonlinear
action of the Lorentz transformations that
incorporates the deformation of boosts. Various
proposals have appeared in the literature for the
associated realization in position space. While some
are based on noncommutative geometries, others respect
the compatibility of the spacetime coordinates. Among
the latter, there exist several proposals that invoke
in different ways the completion of the Lorentz
transformations into canonical ones in phase space. In
this paper, the relationship between all these
canonical proposals is clarified, showing that in fact
they are equivalent. The generalized uncertainty
principles emerging from these canonical realizations
are also discussed in detail, studying the possibility
of reaching regimes where the behavior of suitable
position and momentum variables is classical, and
explaining how one can reconstruct a canonical
realization of doubly special relativity starting just
from a basic set of commutators. In addition, the
extension to general relativity is considered,
investigating the kind of gravity's rainbow that
arises from this canonical realization and comparing
it with the gravity's rainbow formalism put forward by
Magueijo and Smolin, which was obtained from a
commutative but noncanonical realization in position
space.

\vskip 3mm \noindent
\end{abstract}

\maketitle
\newpage
\renewcommand{\thefootnote}{\alph{footnote}}

\section{Introduction}
\setcounter{equation}{0}

Most approaches to quantum gravity support the
existence of a fundamental scale (expected to be of
Planck order) that would act as a threshold for the
inset of quantum effects, so that beyond it the
classical picture of spacetime should be replaced by a
genuine quantum description \cite{Luis,Pad}. For
instance, in loop quantum gravity the classical smooth
geometry of the spacetime is replaced by a discrete
quantum geometry at small scales, with a minimal
nonzero eigenvalue for the area operator that is
indeed of the Planck order \cite{arearov,ash,thierev}.
Similarly, different analyses of scattering processes
in string theory suggest the existence of a minimal
length \cite{GM,ACV,Yon,GKP}. The possible emergence
of this fundamental scale seems to pose a problem if
Lorentz symmetry has to be approximately valid (at
least in some asymptotic regions) in the effective
theory that describes the flat spacetime limit of
quantum gravity. Since, in general, lengths are not
invariant under Lorentz transformations, different
inertial observers would disagree on the value of the
scale at which the quantum discreteness of spacetime
begins to manifest itself. In this sense, the
existence of a fundamental scale would contradict the
relativity principle of inertial frames.

This apparent inconsistency may nonetheless be
overcome by modifying special relativity so as to
preserve an additional scale, apart from the one
provided by the speed of light
\cite{DSRi,DSRii,DSR1,DSR1b,DSR2,DSR2b,DSR12}. Since
the action of the Lorentz transformations must now
leave invariant two dimensionful quantities, the
family of theories that incorporate this suggestion is
generically called Doubly Special Relativity (DSR)
\cite{DSRi}. DSR theories are usually formulated in
momentum space because they lead to modified
dispersion relations \cite{hossenfelder}, which imply
interesting observational consequences
\cite{DSRi,phenomen,phenomena,phenomenb,phenomenc,phenomen2,
phenomen3}. The new scale is an energy and/or momentum
scale, and its invariance is possible only because the
action of the Lorentz group becomes nonlinear in
momentum space.

In order to determine the corresponding action of the
Lorentz group in position space, various proposals
have been discussed in the literature
\cite{kappa,position,mignemi,hinterleitner,ours}. A
class of proposals is based on the use of
noncommutative geometries. Namely, they introduce
spacetime coordinates that are not mutually compatible
in the quantum theory, as it happens e.g. in the case
of $\kappa$-deformed Minkowski spacetime
\cite{DSR12,kappa}. Nevertheless, it is perfectly
feasible to implement DSR theories in position space
while keeping the commutativity of the spacetime
geometry
\cite{hossenfelder,position,mignemi,hinterleitner,ours,rainbow,
cortesgamboa,ours2}. For example, Kimberly, Magueijo
and Medeiros \cite{position} proposed a realization
that preserves the Lorentz invariance of the
contraction between energy-momentum and position. This
requirement ensures that free field theories based
upon DSR admit planewave solutions \cite{position}.
Such a position space realization was later
reformulated and generalized to the framework of
general relativity by Magueijo and Smolin, including
the effect of curvature \cite{rainbow}. The resulting
formalism describes geometries that depend explicitly
on the energy-momentum of the particle that probes
them (in the language of the renormalization group,
geometry ``runs'' \cite{rainbow}). For this reason,
this proposal is called gravity's rainbow.

There exists another class of proposals that, while
maintaining the use of a commutative geometry for the
description of the spacetime, are based (in a way or
another) on a canonical implementation of the Lorentz
transformations rather than on the planewaves
requirement of Ref. \cite{position}. Probably, the
first of these proposals was made by Mignemi
\cite{mignemi}, who demanded covariance on phase space
when studying the transformation law for commuting
coordinates in DSR. Nonetheless, instead of pursuing
this alternative, he finally focused his study on
noncommutative geometries. Another proposal for a
canonical implementation appeared in a work by the
authors \cite{ours}. This proposal employed (the
existence of) a nonlinear map that transforms the
physical energy-momentum into an auxiliary one on
which the Lorentz action becomes linear \cite{judes}.
The realization of DSR in position space was deduced
by completing this map into a canonical
transformation, so that the symplectic form on phase
space remains invariant \cite{ours}. Similarly,
Hinterleitner \cite{hinterleitner} proposed a
prescription for DSR in position space such that the
complete Lorentz transformation on phase space turned
out to be canonical. On the other hand, Cort\'es and
Gamboa \cite{cortesgamboa} analyzed the modification
of quantum commutators in DSR under the assumption
that one may adopt commuting spacetime coordinates
corresponding to the generators of translations in the
auxiliary energy-momentum, so that they form
altogether a canonical set.

Although all these proposals were put forward
independently, it is clear that they are interrelated.
For instance, Hossenfelder pointed out recently
\cite{hossenfelder} that it is possible to use a
generalized uncertainty principle to find the
nonlinear map between the physical and the auxiliary
energy-momentum. This map, together with the results
of Ref. \cite{judes}, suffices to fix the DSR theory.
The main aim of the present work is to discuss in
detail the connection between the different canonical
realizations suggested for DSR. In spite of the
confusion that seems to exist in the literature, we
will prove that they are actually equivalent.
Moreover, extending Hossenfelder's analysis, we will
study in detail the generalized uncertainty principles
associated with this kind of realizations. In
particular, we will provide an explicit construction
of the bijective map between DSR theories and sets of
modified commutators between physical (or auxiliary)
energy-momentum and auxiliary (or respectively
physical) position variables. We will also investigate
the possibility that there exist limiting regimes in
which these mixed (auxiliary and physical) elementary
phase space variables reach a classical behavior.
Finally, following exactly the line of reasoning
defended by Magueijo and Smolin \cite{rainbow} to
extend the DSR realization in position space to
general relativity, we will discuss the gravity's
rainbow formalism that arises from a canonical
realization. We will not consider here other more
tentative suggestions for the extension of DSR to
general relativity, like e.g. that explored in Ref.
\cite{hinterleitner}.

The organization of the rest of the paper is as
follows. In Sec. II we briefly review some basic
aspects of DSR theories formulated in momentum space.
We prove in Sec. III that the different canonical
realizations of DSR in position space that have
appeared in the literature are in fact equivalent.
Section IV deals with the modified quantum commutators
that correspond to canonical realizations of DSR. In
Sec. V we summarize the proposal for a gravity's
rainbow and compare the formalism put forward by
Magueijo and Smolin with the corresponding counterpart
obtained from a canonical realization of DSR. As an
important example, we consider in Sec. VI the
Schwarzschild solution to the modified Einstein
equations obtained in these two types of gravity's
rainbow. Finally, Sec. VII contains the conclusions.

\section{DSR in Momentum Space}
\setcounter{equation}{0}

In DSR, the nonlinear action of the Lorentz
transformations in momentum space can be easily
captured in a nonlinear invertible map $U$ between the
physical energy-momentum $P_{a}:=(-E,p_i)$
$^{\footnotemark[1]}$ \footnotetext[1]{Lowercase Latin
letters from the beginning and the middle of the
alphabet denote Lorentz and flat spatial indices,
respectively.} and an auxiliary energy-momentum that
transforms like the usual momentum variables in
standard special relativity,
$\Pi_{a}:=(-\epsilon,\Pi_i)$ \cite{judes}.
Representing the conventional linear action of the
Lorentz group by $\mathcal{L}$, the nonlinear Lorentz
transformations are given by \cite{DSR2b,judes}
\begin{equation}\label{lorentz}L_a(P_b) = \left[U^{-1}\circ
\mathcal{L}\circ U\right]_a(P_b),\end{equation} where
the symbol $\circ$ denotes composition. Demanding that
the standard action of rotations is not modified, the
map $U$ is totally determined by two scalar functions
$g$ and $f$ \cite{DSR2b,kappa}. Using a notation
similar to that of Refs. \cite{ours,ours2}, $U$ can be
expressed $^{\footnotemark[2]}$ \footnotetext[2]{Our
definition of the functions $f$ and $g$ coincides with
that introduced in Refs. \cite{ours,ours2}, but
differs in general from other conventions found in the
literature. Although our convention simplifies the
calculations, when comparing our results with those of
other works it is important to take into account this
possible discrepancy.}
\begin{equation}\label{momenta}
P_a = U_a^{-1}(\Pi_b) \Rightarrow \left\{ \begin{array}{l}
E = g(\epsilon,\Pi),\\
p_i = f(\epsilon,\Pi)\frac{\Pi_i}{\Pi}.
\end{array} \right.\end{equation}
Here, $\Pi$ is the (Euclidean) norm of the auxiliary
momentum $\Pi_i$. Each admissible choice of functions
$f$ and $g$ leads to a different DSR theory. We
remember that the map $U^{-1}$ must be invertible in
its range. In addition, in order to recover standard
special relativity, $U^{-1}$ must reduce to the
identity (i.e., $g\approx\epsilon$, $f\approx\Pi$) in
the region of negligible energies and momenta compared
to the scale of the DSR theory. This scale corresponds
to the limit of the functions $f$ and/or $g$ when the
auxiliary energy-momentum reaches infinity (on the
mass shell $\epsilon^2-\Pi^2=\mu^2$, where $\mu$ is
the Casimir invariant of the auxiliary momentum
space). So, at least one of the two considered
functions must have a finite limit.

\section{Canonical Implementation of DSR in Position Space}
\setcounter{equation}{0}

For the realization of DSR in position space we will
focus on a family of proposals that, in the context of
commutative geometry, are based on a canonical
implementation of DSR
\cite{mignemi,hinterleitner,ours,ours2,cortesgamboa,hossenfelder}.

Let us start by considering the proposal introduced by
Mignemi in Ref. \cite{mignemi}. Mignemi investigated
the Hamiltonian formalism for a free particle with
physical energy-momentum $P_a$ and associated
commuting spacetime coordinates $x^a$. From Eq.
(\ref{lorentz}), the nonlinear action of the Lorentz
transformations in momentum space is
\begin{equation}\label{momentransfor}
P_a\rightarrow P'_{a}=L_a(P_b).\end{equation} Here, we
have conveniently adapted all expressions to our
notation $^{\footnotemark[3]}$
\footnotetext[3]{Mignemi denotes the action of the
deformed boosts by $W$, rather than using the symbol
$L$ for all Lorentz transformations in DSR. In
addition, he calls $p_a$ the physical energy-momentum
and $q^a$ the spacetime coordinates conjugate to it,
whereas we denote these physical variables by $P_a$
and $x^a$, respectively \cite{ours,ours2}. We reserve
the notation $q^a$ for the auxiliary spacetime
coordinates conjugate to the auxiliary energy-momentum
$\Pi_a$.}. Mignemi identifies then the position space
as the cotangent one to the momentum space. He also
introduces implicitly the demand that the origin of
the spacetime coordinates remain invariant under all
Lorentz transformations. Appealing to covariance, he
deduces the following transformation of coordinates
\cite{mignemi}:
\begin{equation}\label{mignemi}
x^a\rightarrow x'^{a}=\left(\frac{\partial
P'_{a}}{\partial P_b}
\right)^{-1}x^b:=\bar{L}^a(x^b,P_c).\end{equation}
Note that, from Eq. (\ref{momentransfor}), $\partial
P'_{a}/\partial P_b=\partial L_{a}/\partial P_b$. The
above transformation in position space is linear in
the coordinates $x^a$, but depends nontrivially on the
physical energy-momentum.

Soon after Mignemi's proposal, Hinterleitner
\cite{hinterleitner} suggested independently that the
realization of DSR in position space can be determined
by completing the Lorentz transformations into
canonical ones on phase space. Although he restricted
his analysis to two-dimensional spacetimes for
simplicity, we will reproduce his arguments in four
dimensions. He considered canonical transformations
between phase space coordinates in two different
inertial frames, $(x^a,P_a)$ and $(x'^{a},P'_{a})$,
imposing that their restriction to momentum space
coincided with the corresponding Lorentz
transformations in DSR. The canonical transformation
is then totally fixed if one requires that its action
on the spacetime coordinates be homogeneous (i.e. the
origin be left invariant). Calling $P_a=P_a(P'_b)$, it
is straightforward to conclude that $x'^{a}=(\partial
P_b/\partial P'_{a})x^b$, so that (employing the
inverse function theorem) one arrives precisely at
transformation (\ref{mignemi}). Thereby, we see that
Mignemi's and Hinterleitner's proposals are in fact
the same. Let us note that, in spite of the fact that
the two proposals have appeared separately in the
literature, their equivalence is actually obvious once
the condition of covariance is understood as the
invariance of the symplectic form ${\bf d}x^a\wedge
{\bf d}P_a$, which in turn implies that the considered
transformations must be canonical.

On the other hand, in Ref. \cite{ours} we proposed to
find the realization of DSR in position space by
demanding that the extension of the nonlinear map
(\ref{momenta}) to phase space preserve the form ${\bf
d} q^a \wedge {\bf d} \Pi_a$. Remember that $q^a$ are
the auxiliary spacetime coordinates, canonically
conjugate to $\Pi_a$ \cite{ours,ours2}. The above
requirement assigns to the system new, modified
spacetime coordinates $x^a$ that are conjugate to the
physical energy-momentum $P_a$, so that the relation
between $(x^a,P_a)$ and $(q^a,\Pi_a)$ is a canonical
transformation. Demanding that the origins of the two
different sets of spacetime coordinates coincide, one
then gets
\begin{equation}\label{linear}q^a=\frac{\partial P_b}{\partial
\Pi_a}\,x^b.\end{equation} With this relationship
between auxiliary and physical coordinates one can
fully determine the nonlinear action of the Lorentz
group both in position and momentum spaces. It is
given by
\begin{equation}\label{lorentzext}\tilde{L}(x^a,P_a)=[
\tilde{U}^{-1}\circ\tilde{\mathcal{L}}\circ
\tilde{U}](x^a,P_a).\end{equation} Here,
$\tilde{\mathcal{L}}$ represents the standard linear
action of the Lorentz group on phase space, while
$\tilde{U}$ is the canonical map between physical and
auxiliary variables determined by $U$ and Eq.
(\ref{linear}).

Since $\tilde{U}$ is a canonical transformation, so is
its inverse, $\tilde{U}^{-1}.$ Moreover, the standard
Lorentz action $\tilde{\mathcal{L}}$ corresponds as
well to a canonical transformation, which maps
auxiliary variables from an inertial frame to another
one. Hence the composition
$\tilde{L}=\tilde{U}^{-1}\circ\tilde{\mathcal{L}}\circ
\tilde{U}$ provides a canonical transformation. In
addition, its restriction to momentum space reproduces
Eq. (\ref{lorentz}). Thus, $\tilde{L}$ can be regarded
as the canonical extension to phase space of the DSR
transformation of the energy-momentum. Besides,
$\tilde{L}$ leaves the origin of the spacetime
coordinates invariant (because so do $\tilde{U}$ and
$\tilde{\mathcal{L}}$). Therefore, our proposal leads
exactly to the same Lorentz transformation on physical
phase space that was put forward by Hinterleitner. In
this way, our canonical realization of DSR turns out
to be equivalent to those of Mignemi and
Hinterleitner.

The main distinction between our proposal and the ones
suggested by Mignemi and Hinterleitner is that the
latter introduce directly the canonical transformation
between physical variables on phase space as measured
by two different observers, related by means of a
nonlinear Lorentz transformation, whereas our
realization of DSR relies on the canonical map between
auxiliary and physical variables. In other words,
Mignemi and Hinterleitner provide the nonlinear map
$\tilde{L}$ appearing in Eq. (\ref{lorentzext}), while
our proposal gives the map $\tilde{U}$.

Still in the context of commutative geometry, Cort\'es
and Gamboa \cite{cortesgamboa} analyzed the
modification of the quantum commutation relations in
the framework of DSR. They considered spacetime
coordinates interpretable as the generators of
translations in the auxiliary energy and momentum
variables $(-\epsilon,\Pi_i)$. These spacetime
coordinates are therefore canonically conjugate to the
auxiliary energy-momentum and, according to our
notation (which differs from that of Ref.
\cite{cortesgamboa}), they can be identified with
$q^a$. Similarly, one can consider spacetime
coordinates that generate the translations in the
physical energy-momentum $(-E,p_i)$. These are what we
have called the physical spacetime coordinates $x^a$.
The relation between $(q^a,\Pi_a)$ and $(x^a,P_a)$ is
obviously a canonical transformation. In particular,
one can then straightforwardly assign commutators to
the complete set of mixed (i.e. half auxiliary and
half physical) phase space variables $(q^a,P_a)$ by
multiplying their Poisson brackets just by an
imaginary factor $i$ (with the Planck constant $\hbar$
set equal to the unity). In this way, one recovers the
result of Cort\'es and Gamboa, i.e.
\begin{equation}\label{cortesgamboa}[\hat{q}^a,\hat{P}_b]=i
\widehat{\frac{\partial P_b}{\partial
\Pi_a}},\end{equation} the rest of commutators
vanishing. Finally, remember that $\partial
P_b/\partial \Pi_a=\partial U^{-1}_b/\partial \Pi_a$
from Eq. (\ref{momenta}).

In Ref. \cite{hossenfelder} Hossenfelder adopts a very
similar viewpoint. The wave vector $k_a$ introduced in
that work is what we have generically referred to as
physical energy-momentum, $P_a$, and is canonically
conjugate to the physical coordinates $x^a$. This wave
vector is related to the auxiliary energy-momentum via
the nonlinear map $U$. The commutators considered by
Hossenfelder are those corresponding to the mixed
phase space variables $(x^a,\Pi_a)$. Regarded as
fundamental commutators, obtained directly from the
Poisson brackets, they are then given by
\begin{equation}\label{hossenfelder}
[\hat{x}^a,\hat{\Pi}_b]=i \widehat{\frac{\partial \Pi_b}{\partial
P_a}},\end{equation} with $\partial \Pi_b/\partial P_a=\partial
U_b/\partial P_a$.

In conclusion, although the literature contains
several proposals for the canonical realization of DSR
in phase space, we have seen that they are indeed
equivalent, leading to the same transformations of the
spacetime coordinates. We expect that our discussion
helps to clarify this issue definitively.

\section{Modified Commutation Relations}
\setcounter{equation}{0}

At the end of the previous section we have shown that
a canonical realization of DSR yields modified
commutation relations if one chooses as elementary
phase space variables any of the two mixed sets
$(q^a,P_a)$ or $(x^a,\Pi_a)$ (formed by both auxiliary
and physical variables). We want to discuss now the
different behavior of the generalized uncertainty
principles obtained with these two sets. We will also
provide a specific recipe to obtain the map $U$ that
determines the DSR theory starting from the modified
basic commutators.

\subsection{Fundamental commutators}

We first regard as elementary variables the set formed
by the auxiliary spacetime coordinates and the
physical energy-momentum, $(q^a,P_a)$, i.e. we adopt
the choice made in Ref. \cite{cortesgamboa}. From Eqs.
(\ref{momenta}) and (\ref{cortesgamboa}), we obtain
the following nonvanishing commutators:
\begin{eqnarray}\label{brackets}
[\hat{q}^{0},\hat{E}]&=&-i\widehat{\frac{\partial g}{\partial
\epsilon}},\quad \quad\quad\quad\;\;
[\hat{q}^{0},\hat{p}_{i}]=-i\widehat{\frac{\partial
f}{\partial\epsilon}}\widehat{\frac{\Pi_i}{\Pi}}, \\
 \label{brack}
[\hat{q}^{i},\hat{E}] &=&i\widehat{\frac{\partial
g}{\partial
\Pi}}\widehat{\frac{\Pi^i}{\Pi}},\quad\quad\quad\quad
[\hat{q}^{i},\hat{p}_{j}]=i\widehat{\frac{\partial
f}{\partial\Pi}}\widehat{\frac{\Pi^i}{\Pi}}\widehat{\frac{\Pi_j}
{\Pi}}+i\widehat{\frac{f}{\Pi}}\left(\delta^i_j-
\widehat{\frac{\Pi^i}{\Pi}}\widehat{\frac{\Pi_j}{\Pi}}\right).
\nonumber\end{eqnarray} We can view these relations as
a generalized uncertainty principle, noticing that the
usual Heisenberg relations are recovered for energies
and momenta well below the DSR (Planck) scale, because
$g\approx\epsilon$ and $f\approx\Pi$ in this limit. It
is also worth pointing out that all the operators on
the right-hand side of these expressions commute,
because they are determined only by functions of the
energy-momentum, with no contribution from the
position variables $q^a$. In terms of these
fundamental commutators, we can also define the
commutator of other phase space functions.

In order to simplify our expressions and make more
clear the difference between this choice of
fundamental commutators and the one considered by
Hossenfelder, we will restrict our attention in this
subsection to a subfamily of DSR theories in which the
physical energy is just a function of the auxiliary
one, with no dependence on the auxiliary momentum,
namely $E=g(\epsilon)$. These theories have received
special attention in the literature
\cite{cortesgamboa,hossenfelder,rainbow,ours3}. As an
immediate consequence of the fact that  $(\partial
g/\partial \Pi)=0$, the commutator
$[\hat{q}^i,\hat{E}]$ becomes equal to zero. Moreover,
let us consider DSR theories of the so-called DSR2 and
DSR3 types, whose physical energy is bounded from
above. We remember that DSR3 theories have a bounded
physical energy but an unbounded physical momentum,
whereas DSR2 theories possess an invariant scale both
in the physical energy and momentum. In these types of
theories, one has that $(\partial
g/\partial\epsilon)\longrightarrow 0$ when $\epsilon$
approaches infinity and hence the commutator
$[\hat{q}^0,\hat{E}]$ vanishes for infinitely large
auxiliary energy.

On the other hand, the physical momentum is bounded
from above not only in theories of the DSR2 type, but
also in the so-called DSR1 theories, whose physical
energy is nonetheless unbounded. In a theory of the
DSR1 or DSR2 type, the partial derivatives $\partial
f/\partial \epsilon$ and $\partial f/\partial \Pi$ may
both vanish as $\epsilon$ and $\Pi$ tend to infinity
(on the mass shell). We note, however, that while this
vanishing is allowed, it is not a necessary
consequence of the boundedness of the physical
momentum (on the mass shell), because the function $f$
depends on both $\epsilon$ and $\Pi$. If $(\partial
f/\partial \epsilon)\rightarrow 0$ when the auxiliary
energy-momentum approaches infinity, the commutator
$[\hat{q}^0,\hat{p}_i]$ becomes zero in that limit.
Similarly, if $(\partial f/\partial\Pi)\rightarrow 0$,
the commutators $[\hat{q}^i,\hat{p}_j]$ are negligible
when the auxiliary energy-momentum gets large, because
for theories of the DSR1 or DSR2 type one also has
that $(f/\Pi)\longrightarrow 0$ at infinity. Thus, the
operators representing the auxiliary spatial
coordinates would commute with those representing the
physical momentum.

We therefore conclude that all the modified phase
space commutators may vanish for DSR2 theories when
the auxiliary energy-momentum tends to infinity (on
the mass shell). It is then possible to find DSR
theories in which the behavior of the phase space
variables $(q^a,P_a)$ becomes classical when one
approaches the region of invariant energy and momentum
scales.

Let us now analyze the choice of fundamental
commutators made by Hossenfelder \cite{hossenfelder},
which correspond to a set of elementary phase space
variables formed by the physical spacetime coordinates
and the auxiliary energy-momentum. With the
restriction to the subfamily of DSR theories with
$E=g(\epsilon)$, we obtain
\begin{eqnarray}\label{brackhoss}
[\hat{x}^0,\hat{\epsilon}]&=&-i \widehat{\frac{1}{\partial_\epsilon
g}},\quad \quad \quad [\hat{x}^0,\hat{\Pi}_i]=i
\widehat{\frac{1}{\partial_\epsilon
g}}\widehat{\frac{1}{\partial_{\Pi} f}}\widehat{\frac{\partial
f}{\partial \epsilon}}\widehat{\frac{\Pi_i}
{\Pi}},\\
\label{brahos}[\hat{x}^i,\hat{\epsilon}]&=&0,
\quad\quad\quad\quad\quad\;[\hat{x}^i,\hat{\Pi}_j] =i
\widehat{\frac{1}{\partial_{\Pi}
f}}\widehat{\frac{\Pi^i}{\Pi}}
\widehat{\frac{\Pi_j}{\Pi}}+i\widehat{\frac{\Pi}{f}}
\left(\delta^i_j-
\widehat{\frac{\Pi^i}{\Pi}}\widehat{\frac{\Pi_j}{\Pi}}\right).
\nonumber\end{eqnarray} In order to make more compact
our expressions, we have employed a notation of the
type $\partial_\epsilon g:=\partial g/\partial
\epsilon$ for partial derivatives that appear in
denominators.

As we have commented above, $(\partial
g/\partial\epsilon) \longrightarrow0$ when the
auxiliary energy approaches infinity in DSR2 and DSR3
theories. Thus, $[\hat{x}^0,\hat{\epsilon}]$ explodes
in this limit and there is a total lack of resolution
for simultaneous measurements of $x^0$ and $\epsilon$.
On the other hand, for a generic DSR theory,
$[\hat{x}^0,\hat{\Pi}_i]$ may either vanish or tend to
infinity depending just on the functional form of the
partial derivatives of $f$ and $g$, with no a priori
restrictions. Finally, for theories of the DSR1 and
DSR2 type, the ratio $\Pi/f$ (and possibly
$1/\partial_\Pi f$ as well) tends to infinity for
large auxiliary energy-momenta (on the mass shell).
So, for these theories, $[\hat{x}^i,\hat{\Pi}_j]$
diverges. We hence see that all the nontrivial
fundamental commutators may diverge for certain DSR2
theories, with a complete loss of resolution when one
reaches the regime of the invariant (physical) energy
and momentum scales. This result strongly contrasts
with the conclusions obtained for the alternative
choice of modified commutators (\ref{brackets}) that
we have analyzed previously.

One can also study the behavior of the commutators
when the energy of the system is restricted to be a
certain function of the momentum. This interdependence
makes the expressions of the commutators change ``on
shell''. After the reduction to energy surfaces, and
for example with the choice of modified commutators
(\ref{brackets}) made by Cort\'es and Gamboa, one
obtains the generalized uncertainty principle:
\begin{equation} [\hat{q}^i,\hat{p}_j]=i\widehat{\frac{df}{d\Pi}}
\widehat{\frac{\Pi^i}{\Pi}}\widehat{\frac{\Pi_j}{\Pi}}+i
\widehat{\frac{f}{\Pi}}\left(\delta^i_j-
\widehat{\frac{\Pi^i}{\Pi}}\widehat{\frac{\Pi_j}{\Pi}}\right).
\end{equation}
This expression is similar to the last one in Eq.
(\ref{brackets}), but the partial derivative of $f$
with respect to $\Pi$ has been replaced by a total
derivative. This replacement simplifies the analysis
of the emergence of a classical behavior. For DSR1 and
DSR2 theories, it is straightforward to see that the
ratio $f/\Pi$ and the total derivative $df/d\Pi$ tend
to zero when the auxiliary momentum approaches
infinity. Therefore, the commutators of the operators
that represent the auxiliary (spatial) position
variables and the physical momentum vanish in this
limit, which corresponds to the regime in which the
physical momentum reaches the value of the invariant
scale.

Although it may seem initially counterintuitive, the
emergence of a regime with this kind of classical
behavior poses in fact no inconsistency. If one
measures the auxiliary position with infinite
resolution then, according to the standard Heisenberg
principle, one must have a complete uncertainty in its
conjugate momentum, which is the auxiliary one.
However, since DSR1 and DSR2 theories map infinite
auxiliary momenta into a finite physical momentum
scale, the uncertainty in the latter of these momenta
can be kept under control, yielding a vanishing
product with the uncertainty in the auxiliary
position.

\subsection{Derivation of a DSR theory from a
generalized uncertainty principle}

To conclude this section, we will derive explicit
formulas for the nonlinear map $U$ that characterizes
the DSR theory using a set of modified commutators.
These formulas, together with Eq. (\ref{brackets}) or
Eq. (\ref{brackhoss}), prove that the relevant
information present in a canonical realization of a
DSR theory and in a generalized uncertainty principle
is equivalent. We will study explicitly the case of
the fundamental commutators $[\hat{q}^a,\hat{P}_b]$.
The analysis can be straightforwardly extended to the
alternative case of the commutators
$[\hat{x}^a,\hat{\Pi}_b]$.

Note that the functions $f$ and $g$ that determine the
DSR theory, and that in a generic situation depend on
both $\epsilon$ and $\Pi$, appear in the expressions
(\ref{brackets}) as partial derivatives. Therefore,
given a generalized uncertainty principle of the form
(\ref{brackets}), we will obtain the map $U$ (or
equivalently $U^{-1}$) by mere integration. The
reconstruction of an arbitrary function
$z(\epsilon,\Pi)$ from its partial derivatives
$u:=\partial z/\partial \epsilon$ and $v:=\partial
z/\partial \Pi$ is immediate (assuming the
integrability condition $\partial u/\partial
\Pi=\partial v/\partial \epsilon$):
\begin{equation}z(\epsilon,\Pi)=
\int_0^\Pi v(\epsilon,\tilde{\Pi})\,d\tilde{\Pi}+\int_0^\epsilon
u(\tilde{\epsilon},0)\,d\tilde{\epsilon}.
\end{equation}
The function $z$ represents either $f$ or $g$, and we
have employed that $f(0,0)=g(0,0)=0$ to fix an
additive integration constant. It then finally
suffices to realize that
\begin{eqnarray}
\frac{\partial
g}{\partial\epsilon}&=&-\{q^0,E\},\quad\quad\quad\quad\quad
\frac{\partial g}{\partial\Pi}=\{q^i,E\}\frac{\Pi^i}{\Pi},\\
\frac{\partial f}{\partial\epsilon}&=&
-\{q^0,P_i\}\frac{\Pi^i}{\Pi},\quad\quad\quad\;\;
\frac{\partial
f}{\partial\Pi}=\{q^i,P_j\}\frac{\Pi_i}{\Pi}\frac{\Pi^j}{\Pi},
\nonumber\end{eqnarray} where the above Poisson
brackets $\{\cdot,\cdot\}$ are the straight classical
counterpart of the fundamental commutators
(\ref{brackets}) (namely, these brackets are obtained
by replacing energy-momentum operators --that commute
among themselves-- by classical variables, and
removing an imaginary factor $i$).

\section{Gravity's Rainbow}
\setcounter{equation}{0}

As we commented in the Introduction, Magueijo and
Smolin put forward a generalization of DSR that
attempts to incorporate the effect of curvature and,
therefore, provide an extension to general relativity
\cite{rainbow}. This generalization is called
gravity's rainbow and rests on a realization of DSR in
position space that is compatible with the
commutativity of the geometry. The realization
selected by Magueijo and Smolin is not a canonical
one; instead, it is constructed from a specialization
of the requirement (introduced by Kimberly, Magueijo
and Medeiros \cite{position}) that free field theories
in DSR continue to admit planewave solutions. This
specialization can be formulated as the invariance
under Lorentz transformations of the contraction
between the energy-momentum and an infinitesimal
spacetime displacement \cite{rainbow}. In our
notation, this condition can be expressed in the form
$\Pi_a{\bf d}q^a =P_a{\bf d}\check{x}^a.$ We have
denoted the physical coordinates by $\check{x}^a$ to
emphasize that their expressions in terms of auxiliary
phase space variables will now differ from those
obtained with a canonical realization of DSR.

From the above requirement, Magueijo and Smolin
derived an energy dependent set of orthonormal frame
fields. The expressions of the one-forms that they
obtained depend nontrivially on the auxiliary
energy-momentum,
\begin{equation}\label{MS}d\check{x}^0=\frac{\epsilon}{g}dq^0,
\quad \quad\quad\quad
d\check{x}^i=\frac{\Pi}{f}dq^i.\end{equation} Owing to
this dependence, a rainbow of metrics emerge in the
formalism, each particle being associated with a
different metric according to its energy-momentum
\cite{rainbow}.

Actually, from Eq. (\ref{linear}) we see that a
canonical realization of DSR leads also to spacetime
displacements that depend explicitly on the energy and
momentum (which can be understood as corresponding to
the probe used by the observer to test the geometry).
In this sense, canonical DSR theories can be extended
as well to a type of gravity's rainbow formalism
\cite{ours,ours2,ours3}. In the following, we restrict
again our attention to the subfamily of DSR theories
with $E=g(\epsilon)$. It is not difficult to see then
that the requirement of invariance of the symplectic
form ${\bf d}q^a\wedge {\bf d}\Pi_a={\bf d}x^a\wedge
{\bf d}P_a$ leads to the following one-forms
\cite{ours}, rather than to those in Eq. (\ref{MS}):
\begin{eqnarray}\label{dx}dx^{0}&=&{\rm det}
U\left(\frac{\partial f}{\partial\Pi}dq^{0}+\frac{\partial
f}{\partial
\epsilon}\frac{\Pi_i}{\Pi}dq^{i}\right), \\
dx^{i}&=&{\rm det}U\left(\frac{\partial
g}{\partial\epsilon}\frac{\Pi^{i}
\Pi_j}{\Pi^2}dq^{j}\right)
+\frac{\Pi}{f}\left(dq^i-\frac{\Pi^{i}\Pi_j}{\Pi^2}dq^{j}\right).
\nonumber\end{eqnarray} Here, ${\rm det}U=1/[(\partial
g/\partial\epsilon)(\partial f/\partial\Pi)]$. The
analysis carried out by Magueijo and Smolin in Ref.
\cite{rainbow} can then be generalized by substituting
relations (\ref{MS}) with Eq. (\ref{dx}), and will not
be repeated here. Instead, in the next section we
consider a particularly interesting example of the
solutions to the modified Einstein equations that
arise in this type of gravity's rainbow formalism,
namely, the generalization of the Schwarzschild
solution.

\section{Modified Schwarzschild Solution}
\setcounter{equation}{0}

Since we want to study spherically symmetric solutions
of the modified Einstein equations, we will impose a
natural symmetry reduction on the physical phase space
(of the test particle that probes the geometry),
restricting the physical momentum to be parallel or
antiparallel to the spatial position. Therefore, for
each given physical momentum, $\varepsilon^{i}
_{jk}p_i dx^j=0$ \cite{ours3}. Here,
$\varepsilon_{ijk}$ is the Levi-Civit\`a symbol. Since
$p_i/p=\Pi_i/\Pi$, this condition can be rewritten as
$dx^i=dx^j \Pi_j\Pi^i/\Pi^2.$ With this reduction, we
find from Eq. (\ref{dx})
\begin{equation}\label{GM}
d\tilde{x}^0_{\pm}=\frac{1}{\partial_\epsilon g}dq^0,
\quad\quad\quad\quad dx^i=\frac{1}{\partial_{\Pi}f} dq^i,
\end{equation}
where we have defined
\begin{equation}
\tilde{x}^0_{\pm}:=x^0\pm
\frac{1}{\partial_{\epsilon}g}\frac{\partial f}{\partial
\epsilon}x,\quad\quad\quad x:=\sqrt{x^i x^i}.
\end{equation}
The sign in the definition of $\tilde{x}^0_{\pm}$
depends on whether $p^i$ and $dx^i$ are parallel or
antiparallel. Note that the new coordinate
$\tilde{x}^0_{\pm}$, although not canonically
conjugate to the energy, differs from the canonical
time $x^0$ only in the inclusion of a radial shift
vector that is constant in spacetime.

Apart from this constant shift, we notice from Eqs.
(\ref{MS}) and (\ref{GM}) that, in the two
realizations of DSR that we are considering (the
canonical one and that of Ref. \cite{position}), the
geometry is affected just by two independent scalings
that are constant in spacetime, but depend on the
auxiliary energy-momentum. They consist in a time
dilation and a conformal transformation of the spatial
components. The associated (spherically symmetric)
gravity's rainbow formalisms can be explored
simultaneously if one adopts the generic notation
$dq^0=G(\epsilon) d\tilde{x}^0_{\pm}$,
$dq^i=F(\epsilon,\Pi) dx^i$, with $\tilde{x}^0_{\pm}$
designating $x^0$ for the case of the gravity's
rainbow elaborated by Magueijo and Smolin. Explicitly
\begin{equation}\label{FGMS} G(\epsilon):=\frac{g(\epsilon)}
{\epsilon},\quad\quad\quad\quad
F(\epsilon,\Pi):=\frac{f(\epsilon,\Pi)}{\Pi},\end{equation}
for Magueijo and Smolin, whereas for a canonical
realization
\begin{equation} \label{FGGM}G(\epsilon):=
\frac{\partial g}{\partial\epsilon}(\epsilon),\quad\quad\quad\quad
F(\epsilon,\Pi):= \frac{\partial f}{\partial
\Pi}(\epsilon,\Pi).\end{equation}

One can now easily derive the modified Schwarzschild
solution following the same line of reasoning that was
explained in Ref. \cite{rainbow}. In terms of the
physical coordinates (called energy independent by
Magueijo and Smolin), we can generically express a
spherically symmetric metric as \cite{rainbow}
\begin{equation}ds^2=-\tilde{A}(x)[G(\epsilon)]^2
(d\tilde{x}^0_{\pm})^2+
[F(\epsilon,\Pi)]^2\left(\tilde{B}(x)dx^2+x^2
d\Omega^2\right).\end{equation} Translated into auxiliary spacetime
coordinates (energy dependent in Magueijo and Smolin's terminology),
we obtain
\begin{equation}ds^2=-A(q|\epsilon,\Pi)(dq^0)^2+B(q|\epsilon,\Pi)
dq^2+q^2d\Omega^2,\end{equation} where \begin{eqnarray}
q&:=&\sqrt{q^i
q^i},\quad\quad\quad\quad\quad\quad q=F(\epsilon,\Pi) x,\\
A(q|\epsilon,\Pi)&:=&\tilde{A}[x(q,\epsilon,\Pi)]
\quad\quad\quad\;\;
B(q|\epsilon,\Pi):=\tilde{B}[x(q,\epsilon,\Pi)],\nonumber
\end{eqnarray}
and the round metric on $S^2$ is not changed.
According to Birkoff's theorem, the functions $A$ and
$B$ must satisfy
$A(q|\epsilon,\Pi)=B(q|\epsilon,\Pi)^{-1}=1-C(\epsilon,\Pi)/q$
for every fixed auxiliary energy-momentum.

Taking into account the relationship between $q$ and
$x$, we see that, in order to recover the independence
of the function $\tilde{A}(x)$ on the energy-momentum,
the ratio $C(\epsilon,\Pi)/ F(\epsilon,\Pi)$ must be a
genuine constant, not only with respect to its
spacetime dependence, but also with respect to
$\epsilon$ and $\Pi$ \cite{rainbow}. On the other
hand, in the regime of low energies and momenta
$F(\epsilon,\Pi)$ approaches the unity and
$C(\epsilon,\Pi)$ coincides with $2M$, where $M$ is
the Schwarzschild mass of general relativity and we
are setting $G$, the effective Newton constant in the
low energy limit, equal to one. Therefore, we must
have $C(\epsilon,\Pi)= 2M F(\epsilon,\Pi)$.

This completely determines the modified Schwarzschild
solution. In physical coordinates, the explicit
expressions in the gravity's rainbow formalisms
proposed by Magueijo and Smolin, on the one hand, and
arising from a canonical realization, on the other
hand, are respectively
\begin{eqnarray}
ds^2&=&-\left(1-\frac{2M}{x}\right)\!\left(\frac{g}
{\epsilon}\right)^2\!(dx^0)^2+
\left(1-\frac{2M}{x}\right)^{-1}\!\left(\frac{f}
{\Pi}\right)^2\!dx^2+ \left(\frac{f}{\Pi}\right)^2\!x^2d\Omega^2;\\
ds^2&=&-\left(1-\frac{2M}{x}\right)\left(\frac{\partial
g}{\partial\epsilon}\right)^2(d\tilde{x}^0_{\pm})^2+
\left(1-\frac{2M}{x}\right)^{-1}
\!\left(\frac{\partial f}{\partial
\Pi}\right)^2dx^2+\left(\frac{\partial f}{\partial
\Pi}\right)^2 x^2 d\Omega^2.\nonumber\end{eqnarray} In
particular, notice that the horizon is placed at the
same location as in general relativity, $x=2M$. The
corresponding notions of null and spatial infinities
and of asymptotic flatness in the context of DSR can
be introduced as explained in Ref. \cite{hackett}.

\section{Summary and Conclusions}

We have studied several proposals for the realization
of DSR in position space that are based on a canonical
implementation in phase space. We have shown that,
although these proposals have been discussed
separately in the literature, they are actually
equivalent. Some of these proposals have been
formulated as the requirement that the action of the
Lorentz transformations on the physical spacetime
coordinates be such that the total transformation in
the resulting phase space be canonical
\cite{mignemi,hinterleitner}. In another case, the
demand of a canonical transformation has been
introduced instead on the nonlinear map that relates
the physical phase space with an auxiliary one where
the Lorentz transformations have the standard linear
action \cite{ours,ours2}. Finally, in other cases the
realization in position space has been encoded in a
generalized uncertainty principle, with modified
commutators that correspond either to the physical
energy-momentum and the spacetime coordinates that
generate translations in the auxiliary energy-momentum
\cite{cortesgamboa} or, alternatively, to the opposite
choice of auxiliary versus physical variables
\cite{hossenfelder}.

We have also analyzed the behavior of those two
different kinds of modified commutation relations. For
the first kind, namely, when one considers auxiliary
spacetime coordinates and physical energy-momentum, we
have shown that all phase space commutators may vanish
in the limit of large auxiliary energy and momentum
(on the mass shell) for certain DSR theories of the
so-called DSR2 type. Therefore, it is in principle
possible to approach a classical regime for the
mentioned elementary phase space variables when the
physical energy-momentum tends to the invariant scale
of the DSR theory. On the contrary, for the same type
of DSR theories, the alternative choice of physical
spacetime coordinates and auxiliary energy-momentum as
elementary variables leads to modified commutators
that generally explode when the system approaches the
invariant energy-momentum scale. In terms of these
elementary variables, there is a complete loss of
resolution in the considered limit. In addition, we
have shown how to construct the nonlinear map $U$ that
determines the DSR theory from the sole knowledge of a
given set of modified phase space commutators. In this
construction, it has been essential that both the
fundamental commutators and certain projections of
them in the momentum direction are not affected by
factor ordering ambiguities, because they are defined
exclusively in terms of functions of the
energy-momentum.

Finally, we have also discussed the possibility of
generalizing the gravity's rainbow formalism put
forward by Magueijo and Smolin in order to find an
extension  to general relativity of the canonical
realization of DSR theories. We have shown that the
generalization is possible and we have compared the
gravity's rainbow that results from the canonical
realization with that elaborated in Ref.
\cite{rainbow} starting from a different proposal for
the realization of DSR in position space
\cite{position}. In both cases, one actually arrives
to a rainbow of metrics that depend nontrivially on
the energy-momentum of the test particles. We have
studied in detail the particular example of the
generalized Schwarzschild solution to the modified
Einstein equations. The main modification with respect
to the standard solution in general relativity
consists in a time dilation and a spatial conformal
transformation, both of them dependent on the
energy-momentum. We have deduced the explicit form of
the dilation and conformal factors in the case of the
canonical realization, showing that they generally
differ from those obtained by Magueijo and Smolin.
This difference may have important consequences for
black hole thermodynamics \cite{ours3}.

\acknowledgments

The authors are grateful to L. Garay and C. Barcel\'o
for conversations. P. G. gratefully acknowledges the
financial support provided by the I3P framework of
CSIC and the European Social Fund. This work was
supported by funds provided by the Spanish MEC Project
No. FIS2005-05736-C03-02.

\end{document}